\documentclass{aa}
\usepackage{graphicx}
\usepackage{amsmath}
\usepackage{txfonts}

\begin{document}

\title{Vacuum ultraviolet photolysis of hydrogenated amorphous carbons.}
\subtitle{III. Diffusion of photo-produced H$_{2}$ as a function of temperature} 
\author{R. Mart\'in-Dom\'enech \inst{1}, E. Dartois \inst{2}, and G.~M. Mu\~noz Caro \inst{1}}

\institute{Centro de Astrobiolog\'{\i}a, INTA-CSIC, Carretera de Ajalvir, km 4, Torrej\'on de Ardoz, 28850 Madrid, Spain\\
\and Institut d'Astrophysique Spatiale, UMR8617, CNRS/Universit\'e Paris-Sud, Universit\'e Paris-Saclay, Universit\'e Paris-Sud, 
F-91405 Orsay, France\\}
\date{Received - , 2016; Accepted - , 2016}

  \abstract
{Hydrogenated amorphous carbon (a-C:H) has been proposed as one of the carbonaceous solids detected in the interstellar medium. 
Energetic processing of the a-C:H particles leads to the dissociation of the C-H bonds and the formation of hydrogen molecules and small 
hydrocarbons. 
Photo-produced H$_{2}$ molecules in the bulk of the dust particles can diffuse out to the gas phase and contribute to the total H$_{2}$ abundance.}
{We have simulated this process in the laboratory with plasma-produced a-C:H and a-C:D analogs under astrophysically relevant conditions 
to investigate the dependence of the diffusion as a function of temperature.} 
{Experimental simulations were performed in a high-vacuum chamber, with complementary experiments carried out in an ultra-high-vacuum 
chamber. 
Plasma-produced a-C:H analogs were UV-irradiated using a microwave-discharged hydrogen flow lamp. 
Molecules diffusing to the gas-phase were detected by a quadrupole mass spectrometer, providing a measurement of the outgoing H$_{2}$ or D$_{2}$ 
flux. 
By comparing the experimental measurements with the expected flux from a one-dimensional diffusion model, a diffusion coefficient $D$ could 
be derived for experiments carried out at different temperatures.} 
{Dependance on the diffusion coefficient $D$ with the temperature followed an Arrhenius-type equation. 
The activation energy for the diffusion process was estimated ($E_{D}$(H$_{2}$) = 1660 $\pm$ 110 K, $E_{D}$(D$_{2}$) = 2090 $\pm$ 90 K), 
as well as the pre-exponential factor 
($D_{0}$(H$_{2}$) = 0.0007$^{+0.0013}_{-0.0004}$ cm$^{2}$ s$^{-1}$, $D_{0}$(D$_{2}$) = 0.0045$^{+0.005}_{-0.0023}$ cm$^{2}$ s$^{-1}$)}
{The strong decrease of the diffusion coefficient at low dust particle temperatures exponentially increases  the diffusion times 
in astrophysical environments. 
Therefore, transient dust heating by cosmic rays needs to be invoked for the release of the photo-produced H$_{2}$ molecules 
in cold PDR regions, where destruction of the aliphatic component in hydrogenated amorphous carbons most probably takes place.}


\maketitle

%
%
\section{Introduction}
\label{intro}
Dust particles in the interstellar medium (ISM) include minerals (silicates and oxides) and carbonaceous matter of various types. 
In the diffuse ISM, carbonaceous solids are observed through both emission and absorption bands in the mid-infrared (mid-IR) 
region of the spectrum.
The so-called aromatic infrared bands (AIBs) are a group of emission bands ubiquitously observed notably at 3.3, 6.2, 7.7, 8.6, and 11.3 $\mu$m,  
generally associated with the infrared fluorescence of polycyclic aromatic hydrocarbons \citep[PAHs, ][]{leger84,allamadola85} 
upon absorption of ultraviolet (UV) photons. 
The observed AIBs spectral variabilities have been classified in three main types (named A, B, and C), 
reflecting the evolution of the carriers in the environment 
\citep{peeters02,vanDied04}, 
which would account for 4-5\% of the total cosmic carbon abundance \citep{draine07}.
 
An absorption band observed at 3.4 $\mu$m toward several lines of sight \citep{soifer76,allen80,mcfadzean89,adamson90,sandford91,pendleton94,
bridger94,imanishi00a,imanishi00b,spoon04,dartois07} 
has been assigned to hydrogenated amorphous carbons (a-C:Hs or HACs),  
that harbor 5-30\% of the cosmic carbon, depending on the assumed carrier \citep[see, e.g., ][]{sandford91}.  
The 3.4 $\mu$m ($\sim$2900 cm$^{-1}$) feature arises from the contribution of the symmetric and asymmetric C-H stretching modes 
of the methyl (-CH$_{3}$) and methylene (-CH$_{2}$-) groups. 
The corresponding bending modes are also observed at 7.25 $\mu$m ($\sim$1380 cm$^{-1}$), and 6.85 $\mu$m ($\sim$1460 cm$^{-1}$), respectively. 
These bands are accompanied by a broad absorption between 6.0 and 6.4 $\mu$m attributed to aromatic and olefinic C=C stretching modes,   
although the carrier is predominantly aliphatic \citep{dartois07}. 
Several laboratory analogs have been proposed to fit these observed IR features \citep{schnaiter98,lee93,furton99,mennella99,mennella03,
dartois05,godard10,godard11}. 
In this paper, we focus on plasma-produced a-C:H analogs \citep[see ][and Section \ref{lab}]{godard10,godard11}. 

The 3.4 $\mu$m absorption band is not detected, though, in the dense ISM. 
Laboratory simulations on the energetic processing of HAC analogs under astrophysically relevant conditions 
have resulted in a decrease of their hydrogen content and, therefore, of the aliphatic C-H spectral features  
\citep[see, e.g., ][and references therein]{mennella03,godard11,alata14,alata15}. 
This energetic processing is driven by the interaction of the hydrogenated amorphous carbon particles with ultraviolet (UV) photons in the 
diffuse ISM \citep[dehydrogenation by cosmic rays is negligible in these regions, see][]{mennella03}
while destruction by cosmic rays (directly, or indirectly through the generated secondary UV field) dominates in the dense ISM. 
The presence of atomic hydrogen allows re-hydrogenation in the diffuse ISM. 
However, this process is inhibited in the dense ISM, possibly leading to a more aromatic carbonaceous solid (amorphous carbons or a-Cs) 
if not fully destroyed at earlier times. 
The destruction/transformation of the aliphatic C-H component most probably takes place in intermediate regions such as translucent clouds or 
photo-dominated regions, where dehydrogenation by both 
UV photons and cosmic rays is still active while hydrogen is in molecular form, thus not allowing re-hydrogenation \citep{godard11}. 

Hydrogen atoms resulting from the rupture of C-H bonds recombine to form H$_{2}$ molecules. 
Molecular hydrogen subsequently diffuses to the surface and is lost from the a-C:H \citep{wild87,moller87,adel89,maree96,godard11,alata14}. 
This process thus constitutes an alternative pathway for H$_{2}$ formation within the bulk of the hydrogenated amorphous carbon particles in 
the ISM, in addition to the previously studied formation on the surface of interstellar solids from physisorbed and/or chemisorbed H atoms  
\citep[see, e.g., ][]{pirronello97,katz99,habart05,cazaux11,gavilan12}. 
Although molecular hydrogen is the main product resulting from the photolysis of a-C:H analogs, small hydrocarbons with up to 
four carbon atoms are also detected \citep{alata14,alata15}. 
Production of these small hydrocarbons in the bulk of carbonaceous solids is proposed as an additional source that may account for the abundance of 
these species in some regions where pure gas-phase models face difficulties in predicting them \citep{pety05,alata15}.  

In this work, we investigate the diffusion of molecular hydrogen through the plasma-produced hydrogenated amorphous carbon analogs 
under astrophysically relevant conditions. 
Deuterated analogs have been preferently used to avoid confusion with background H$_{2}$ during the first experiments. 
We have focused on the variation of the diffusion coefficient as a function of the temperature and, in particular, 
we have estimated $D_{0}$ and $E_{D}$ for the diffusion of H$_{2}$ (D$_{2}$) molecules through the a-C:H (a-C:D) analogs. 
The paper is organized as follows: 
Section \ref{lab} describes the experimental setup and the theoretical models used to evaluate the diffusion coefficient from the experiments. 
Section \ref{results} presents the experimental results, while their astrophysical implications are discussed in Sect. \ref{astro}. 
Finally, conclusions are summarized in Sect. \ref{conclusiones}. 


\section{Methods}
\label{lab}

\subsection{SICAL-X}
\label{sicalx}
The majority of the experiments were carried out using the SICAL-X setup described in previous papers \citep[see, e.g., ][]{alata14} 
at the Institut d'Astrophysique Spatiale. 
The SICAL-X setup consists in a high-vacuum (HV) chamber with a base pressure of about 2 $\times$ 10$^{-8}$ mbar. 
At this pressure, residual H$_{2}$ inside the chamber is not negligible. 
Therefore, deuterated amorphous carbon analogs were preferentially used to study diffusion of in situ photo-produced D$_{2}$ 
within this material. 
The a-C:D analogs produced in a different setup (SICAL-P, see Sect. \ref{sicalp}) on a MgF$_{2}$ substrate 
were introduced in the chamber and cooled down to the working temperature thanks to the combination of a closed-cycle helium 
cryostat and a resistive-type heater. 
The sample temperature was monitored with a thermocouple, reaching a sensitivity of 0.1 K. 

Solid samples were monitored with a Bruker Vertex 80v Fourier transform infrared (FTIR) spectrometer. 
Spectra were collected at a resolution of 1 cm$^{-1}$, covering the spectral region between 7500 and $\sim$1500 cm$^{-1}$ 
(due to the low transmittance of the MgF$_{2}$ substrate below $\sim$1500 cm$^{-1}$).  
The analogs thicknesses $l$ were estimated from the interference pattern (or fringes) in the IR spectra, using the formula

\begin{equation}
 l = 1/(2 \cdot n \cdot \Delta\sigma \cdot cos(\alpha_{IR})), 
 \label{fringes}
\end{equation}

where $n$ is the refractive index of the analogs, estimated to be 1.7 $\pm$ 0.2, $\Delta\sigma$ the fringe spacing, and $\alpha_{IR}$ the 
angle of incidence of the IR beam with the sample normal (45$^{\circ}$ in SICAL-X). 

\begin{figure*}
\includegraphics[width=12cm]{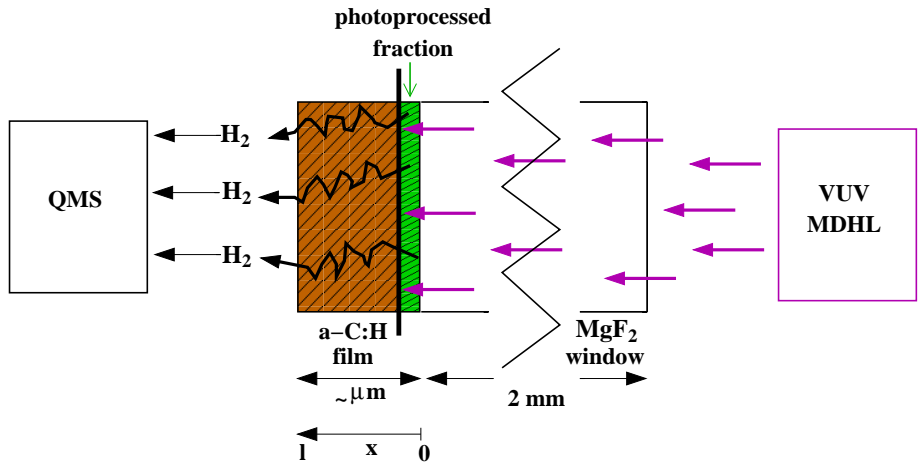}
\centering
\caption{Configuration of the measurement.}
\label{setup}
\end{figure*}

To study the diffusion of D$_{2}$ molecules through the a-C:D analogs, samples were irradiated with vacuum-ultraviolet (VUV) photons,  
leading to the dissociation of C-D bonds and the formation of D$_{2}$ in the analogs. 
Photochemical properties of a-C:H and a-C:D analogs are expected to be similar \citep[see ][]{alata14}. 
UV photons reached the surface of the analogs directly in contact with the MgF$_{2}$ substrate, 
and the photo-produced D$_{2}$ molecules diffused through the samples to the opposite surface, 
subsequently passing into the gas phase (see Fig. \ref{setup}).    
The mean penetration depth for the VUV photons in the a-C:D analogs was approximately 80 nm as measured with VUV-dedicated experiments 
\citep{gavilan15}. 
Since the plasma-produced analogs had a thickness of around 1 $\mu$m, 
the region affected by the VUV photons was negligible compared to the total thickness of the sample, which remained almost unaltered. 
The surface of the a-C:D analogs was around 1 cm$^{2}$, much larger than the thickness of the samples. 
Therefore, only diffusion in the direction orthogonal to the substrate was considered. 

UV irradiation was performed using a microwave-discharged hydrogen flow lamp (MDHL), 
with a VUV-photon flux at the sample position of 2.7 $\times$ 10$^{14}$ photons cm$^{-2}$ s$^{-1}$, 
measured by actinometry \citep{alata14}. 
Hydrogen pressure was set to 0.7 mbar in the lamp. 
The VUV spectrum of the MDHL has been studied in previous papers \citep[see, e.g., ][and references therein]{gustavo14,alata14}. 
An MgF$_{2}$ window was used as the interface between the lamp and the chamber interior. This window, along with the substrate, led to a 
cutoff in the lamp emission at about 115 nm. 
The mean photon energy was 8.6 eV. 
A 20 mm diameter metallic shutter was placed 5 cm away in front of the substrate to prevent the VUV photons from reaching the 
sample when necessary, without turning off the lamp and changing the conditions inside the chamber during the experiments (blank conditions). 

The initial D$_{2}$ concentration throughout the a-C:D analogs was zero. 
Photo-production of D$_{2}$ at the irradiated a-C:D surface in contact with the MgF$_{2}$ substrate established a constant 
D$_{2}$ flux entering the samples, 
since the IR spectra of the analogs did not change after UV irradiation.  
Deuterium molecules that were subsequently diffused through the samples eventually reached the opposite surface, where they inmediately passed into the 
gas-phase. 
Deuterium concentration at the sample surface in contact with the gas phase was thus negligible during the experiments. 
D$_{2}$ molecules in the gas phase were detected by a Quadera 200 quadrupole mass spectrometer (QMS) located at $\sim$10 cm from the samples. 
A heated filament in the QMS produced a stable current of energetic electrons ($\sim$70 eV), which ionized the D$_{2}$ molecules by electron 
bombardment. 
Ions were subsequently detected by a secondary-emission multiplier (SEM) detector. 
The ion current corresponding to the m/z = 4 mass fragment provided a measure of the D$_{2}$ flux through the surface of the a-C:D 
analogs in contact with the gas phase, considering the background level as zero flux. 
Therefore, upon onset of the VUV irradiation, D$_{2}$ concentration throughout the a-C:D analogs, as well as the measured outgoing flux of D$_{2}$ 
molecules, increased to a steady-state value \citep[see][]{alata14}. 
Once the steady state was reached, the shutter was placed in front of the sample, blocking the VUV photons and stopping the D$_{2}$ 
production, thus eliminating the entering D$_{2}$ flux at the sample surface in contact with the substrate. 
As a consequence, the measured m/z = 4 ion current (i.e., the outgoing D$_{2}$ flux) decreased back to the background value.
  
The evolution of the measured D$_{2}$ flux with time during irradiation of the sample (flux increasing to a steady-state value), 
and after irradiation (when D$_{2}$ formation is stopped and the m/z = 4 ion current decreases from the steady-state back to the 
background value) 
depends on the diffusion coefficient $D$ of the D$_{2}$ molecules through the a-C:D analogs. 
The diffusion coefficient can thus be estimated from the one-dimensional diffusion models that provide outgoing flux values that best fit 
the measured ion currents during the experiment (see Sect. \ref{modelo}). 

Since the deuterium molecules were produced in situ in the a-C:D analogs during irradiation of the samples, we could not a priori disentangle 
the different steps taking place in the process: 
rupture of the C-D bonds, diffusion of the D atoms and recombination of two D atoms or direct neighbor D-abstraction 
from a C-D bond to form D$_{2}$ molecules, and the 
diffusion process itself. Therefore, the derived diffusion coefficients should be seen, in principle, as so-called apparent coefficients, describing the 
convolution of all these steps. 
However, 
D$_{2}$ molecules passing into the gas phase were detected very early on, once UV irradiation was established, 
subsequently increasing the observed flux with time (see Figures \ref{irradiationhac}, \ref{irradiationhac2}, and left panel of Fig. \ref{ajuste}, 
in Sect. \ref{results} where the experimental results are presented). This means that  
all processes prior to the diffusion of molecules were probably taking place in a much shorter timescale than the diffusion itself, which 
could be considered the limiting step. 
In particular, Fig. \ref{irradiationhac2} in Sect. \ref{results} shows the symmetry between the measured m/z = 4 ion current during 
irradiation (increasing curve) 
and the decreasing signal observed once the UV beam was blocked after reaching the steady-state (when the diffusion equilibrium is achieved 
and the film is full of D$_{2}$). 
This shows that the measurement is dominated by the diffusion step with respect to the molecule formation timescale. 
Otherwise, if D$_{2}$ formation was the limiting step, we would expect to observe a delay and asymmetry 
in the m/z = 4 signal for the increasing curve owing  to the D$_{2}$ formation limiting steps at the beginning, 
compared to the decreasing curve, when the beam is off and the bulk of the film is full of of previously formed D$_{2}$ molecules. 
Other measured behaviors supported the fact that the D$_{2}$ formation steps occurred at much shorter timescales than the diffusion step  
for the film thicknesses used and at the temperatures we performed the experiments, as explained below.
  
%

The diffusion coefficient $D$ of a diffusing species through a given material is temperature dependent. 
Therefore, consecutive experiments following the above explained protocol were carried out with the same sample at different temperatures, 
allowing the evaluation of the diffusion coefficient as a function of the temperature. 
After every experiment, the a-C:D analogs were warmed up to 250 K with a heating rate of 5 K/min, to evacuate the eventual remaining deuterium from the sample, and set the 
D$_{2}$ concentration back to zero before performing the experiment at a different temperature.  

According to Equation \ref{diffT}
 (see Sect. \ref{modelo}), the diffusion coefficient increases with increasing temperature. 
Therefore, the diffusion time decreases with increasing temperature for a given sample thickness. 
At the same time, diffusion of deuterium through thicker samples takes longer,  
since molecules have to go through a longer distance to reach the analog surface in contact with the gas phase.   
A set of a-C:D analogs with different thicknesses were used to study diffusion in a wide range of temperatures while keeping the duration of 
the experiments within reasonable limits.  
In this way, diffusion at low temperatures (95 K - 140 K) was preferentially probed with thinner samples (0.2 $\mu$m - 2.6 $\mu$m), 
since diffusion times were shorter despite the lower diffusion coefficient; 
while diffusion at high temperatures (110 K - 170 K) was studied with thicker analogs (3.4 $\mu$m - 5.2 $\mu$m). 
When thinner films were irradiated at the same temperature (i.e., the diffusion length was short and bulk diffusion timescale was rapid), 
the molecular D$_{2}$ release was observed immediately after turning the UV lamp on, and stopped immediately after switching it off, 
with no delay that would otherwise indicate a longtime scale limiting the D$_{2}$ formation step 
(meaning with timescales of the order of seconds, which was the QMS scanning time). 
These test measurements did not allow us to measure the bulk diffusion and are thus not shown in Sect. \ref{results}.
In addition, the diffusion timescales changed with the thickness according to what was expected for different films  
at the same temperature (see above), whereas the production rate of D$_{2}$ on one side was confined to the same small thickness ($\sim$80 nm, 
as explained above) for all films, supporting the fact that diffusion was the limiting step.
 
\subsection{ISAC}
\label{isac}
Complementary experiments were carried out with a-C:H analogs (also produced in the SICAL-P setup on MgF$_{2}$ substrates, see Sect. \ref{sicalp}) 
using the ISAC setup \citep{munozcaro10}, 
to evaluate the diffusion coefficient $D$ of H$_{2}$ molecules through hydrogenated amorphous carbon analogs as a function of the temperature, 
Since H$_{2}$ molecules are smaller than D$_{2}$ molecules, diffusion coefficient of the former is expected to be higher than that of the 
latter at a given temperature. 
However, dependence of the diffusion coefficient of H$_{2}$ through the a-C:H analogs as a function of the temperature 
is expected to be similar to that of the diffusion coefficient of D$_{2}$ through a-C:D analogs, 
leading to similar $E_{D}$ values. 
The ISAC setup consists in an ultra-high-vacuum (UHV) chamber with a base pressure of about 4 $\times$ 10$^{-11}$ mbar, three orders of 
magnitude lower than that of the SICAL-X setup (see Sect. \ref{sicalx}), thus enabling us to work with H$_{2}$ instead of D$_{2}$. 
The a-C:H analogs were introduced in the chamber and cooled down to the working temperature, also using  a closed-cycle helium cryostat 
and a resistive-type heater. 
The temperature was controlled thanks to a silicon-diode sensor and a LakeShore Model 331 controller, 
reaching a sensitivity of 0.1 K. As in the SICAL-X setup, solid samples were monitored with a Bruker Vertex 70 FTIR spectrometer. 
Spectra were collected with a spectral resolution of 2 cm$^{-1}$, covering the range between 6000 and $\sim$1500 cm$^{-1}$. 
The angle of incidence of the IR beam with the sample normal was 0$^{\circ}$ in this setup. 

Diffusion of H$_{2}$ molecules through the a-C:H analogs were studied using the experimental protocol described in Sect. \ref{sicalx}. 
The VUV photon flux of the MDHL at the sample position was about 2 $\times$ 10$^{14}$ photons cm$^{-2}$ s$^{-1}$, measured by 
CO$_{2}$ $\to$ CO actinometry \citep{munozcaro10}. 
A Pfeiffer Prisma QMS with a Channeltron detector located at $\sim$17 cm apart from the sample was used to detect the H$_{2}$ molecules in the 
gas phase, which were also ionized by electron bombardment with energetic ($\sim$70 eV) electrons.  
The ion current of the m/z = 2 mass fragment provided a measure of the outgoing H$_{2}$ flux. 
No shutter was present in the ISAC setup, and we had to turn off the lamp after the steady state was reached to stop the production of H$_{2}$ 
molecules. 
Subsequent changes in the background m/z = 2 ion current were taken into account.

\subsection{SICAL-P}
\label{sicalp}
The a-C:H and a-C:D analogs were prepared by a plasma-enhanced vapor chemical deposition (PECVD) method \citep{godard10,godard11}, 
using CH$_{4}$ and CD$_{4}$ as gas precursors, respectively. Radicals and ions of the low pressure plasma produced by radio frequency (RF) 
at 2.45 GHz were deposited on a MgF$_{2}$ substrate. 
The precursor pressure was kept at $\sim$1 mbar in the SICAL-P vacuum chamber, and the power applied to the plasma was set to 100 W 
for all samples. Consequently, structure of all plasma-produced analogs was expected to be similar. 
Deposition time ranged from a few seconds to half an hour, leading to a wide range of sample thicknesses. 
Hydrogenated and deuterated samples were subsequently transferred to the ISAC or the SICAL-X setup, respectively.

\subsection{Theoretical models}
\label{modelo}
Produced H$_{2}$ in the bulk of the a-C:H particles diffuses to the surface and is subsequently released from the solid. 
Diffusion of H$_{2}$ in a particular direction inside the HAC material can be  
described by the Fick's first law: 

\begin{equation}
F(x,t) = - D \cdot \frac{\partial C(x,t)}{\partial x}, 
\label{fick1}
\end{equation}

where $F(x,t)$ is the rate of transfer of H$_{2}$ molecules through unit area of HAC section (i.e., the H$_{2}$ flux), in cm$^{-2}$ s$^{-1}$, 
$D$ is the diffusion coefficient of molecular hydrogen in the hydrogenated amorphous carbon material, in cm$^{2}$ s$^{-1}$, 
and $\tfrac{\partial C(x,t)}{\partial x}$ is the concentration gradient of H$_{2}$ molecules in the particular direction $x$, in cm$^{-4}$. 
The negative sign indicates that diffusion takes place in the direction of decreasing concentration. 
The diffusion coefficient $D$ usually depends on the diffusing species, the material through which it is diffusing, and the temperature.  
The dependence of $D$ with the temperature generally follows an Arrhenius-type equation: 

\begin{equation}
D = D_{0} \cdot e^{\tfrac{-E_{D}}{T}}, 
\label{diffT}
\end{equation}

where the pre-exponential factor $D_{0}$ is the diffusion at infinite temperature, in cm$^{2}$ s$^{-1}$, 
$E_{D}$ is the activation energy for the diffusion process in K, and $T$ the temperature in K. 

As mentioned in Sect. \ref{sicalx}, the diffusion coefficient $D$ can be estimated from one dimensional 
diffusion models that provides outgoing flux values that best fit the measured H$_{2}$ or D$_{2}$ ion currents during the experiments. Diffusion models are particular solutions of the fundamental differential equation of diffusion, 
which is also known as the Fick's second law:  

\begin{equation}
\frac{\partial C(x,t)}{\partial t}=D \cdot \frac{\partial^{2} C(x,t)}{\partial x^{2}}.  
\label{fick2}
\end{equation}

This equation results from combining Equation \ref{fick1} with a mass balance equation applied to a differential 
element of volume of the material through which diffusion is being studied. 

Particular solutions of Equation \ref{fick2} depend on the initial and surface conditions, leading to a series of diffusion models. 
Concentration of H$_{2}$ or D$_{2}$ in our samples during UV-irradiation (i.e., 
when a constant flux $F_{i}$ is established at the irradiated surface ($x = 0$) 
of a sample of thickness $l$, 
the initial H$_{2}$ or D$_{2}$ concentration is zero throughout the sample, 
and concentration at $x=l$ is $C(l,t)=0$ at all times)  
is modelled by the following equation \citep{early78}:

\begin{equation}
\begin{split}
C(x,t) = \frac{F_{i}}{D} (l - x) - 
\frac{8 F_{i} l}{\pi^{2} D} \times \sum_{n = 0}^{\infty} & \frac{(-1)^{n}}{(2n + 1)^{2}} 
\cdot exp[-\frac{(2n+1)^{2} \pi^{2} D t}{4 l^{2}}]\\ 
\cdot & sin\frac{(2n+1) \pi (l - x)}{2l}. 
 \label{csubida1b}
 \end{split}
\end{equation}

Using Eq. \ref{fick1} at $x=l$, the outgoing flux measured by the QMS, according to this model, can be calculated from Equation \ref{csubida1b}:

\begin{equation}
F(l,t) = 1 - \frac{4}{\pi} \times \sum_{n = 0}^{\infty}\frac{(-1)^{n}}{(2n + 1)} 
\cdot exp[-\frac{(2n+1)^{2} \pi^{2} D  t}{4  l^{2}}], 
\label{subida1b}
\end{equation}

normalized to the steady-state value of the outgoing flux ($F(l,\infty) = 1$). 

On the other hand, concentration of H$_{2}$ or D$_{2}$ in the amorphous carbon analogs after UV-irradiation (i.e., 
when no flux is introduced at the surface $x = 0$ of a sample of thickness $l$, 
the initial H$_{2}$ or D$_{2}$ concentration is $C_{0}$ at $x=0$, 
and concentration at $x=l$ is $C(l,t)=0$ at all times) 
is modelled by the following equation \citep{carslaw}:
\begin{equation}
\begin{split}
 C(x,t) = \frac{8 C_{0}}{\pi^{2}} \times \sum_{n = 0}^{\infty} & \frac{1}{(2n + 1)^{2}} 
 \cdot exp[-\frac{(2n+1)^{2} \pi^{2} D t}{4 l^{2}}]\\ 
 \cdot & cos\frac{(2n+1) \pi  x}{2l}.
 \label{cbajada1D}
 \end{split}
\end{equation}

According to this model, the outgoing flux measured by the QMS can be derived from Equation \ref{cbajada1D} using Equation \ref{fick1} 
at $x=l$:

\begin{equation}
F(l,t) = \frac{4}{\pi} \times \sum_{n = 0}^{\infty}\frac{(-1)^{n}}{(2n + 1)} 
\cdot exp[-\frac{(2n+1)^{2} \pi^{2} D t}{4 l^{2}}], 
\label{bajada1D}
\end{equation}

normalized to the initial (steady-state) value ($F(l,0) = 1$). 
We note that Eq. \ref{subida1b} = 1 - Eq. \ref{bajada1D}. 
As shown in Fig. \ref{irradiationhac2}, evolution of the H$_{2}$ or D$_{2}$ ion current after irradiation is symmetric with 
respect to the rise of the signal during irradiation. 

In our case, evolution of the m/z = 2 and m/z = 4 ion currents during the experiments 
was found to be better described by adding a second term in Equations \ref{subida1b} and \ref{bajada1D}, 
introducing an additional diffusion coefficient $D'$. 
This led to

\begin{equation}
\begin{split}
F(l,t) =1 - &p \cdot \frac{4}{\pi} \times \sum_{n = 0}^{\infty}\frac{(-1)^{n}}{(2n + 1)} 
\cdot exp[-\frac{(2n+1)^{2} \pi^{2} D t}{4 l^{2}}]\\
+ &(1 - p) \cdot \frac{4}{\pi} \times \sum_{n = 0}^{\infty}\frac{(-1)^{n}}{(2n + 1)} 
\cdot exp[-\frac{(2n+1)^{2} \pi^{2} D' \cdot t}{4 l^{2}}], 
\label{subida2D}
\end{split}
\end{equation}

and

\begin{equation}
\begin{split}
F(l,t) = & p \cdot \frac{4}{\pi} \times \sum_{n = 0}^{\infty}\frac{(-1)^{n}}{(2n + 1)} 
\cdot exp[-\frac{(2n+1)^{2} \pi^{2} D t}{4 l^{2}}]\\ 
+ & (1-p) \cdot \frac{4}{\pi} \times \sum_{n = 0}^{\infty}\frac{(-1)^{n}}{(2n + 1)} 
\cdot exp[-\frac{(2n+1)^{2} \pi^{2} D' t}{4 l^{2}}], 
\label{bajada2D}
\end{split}
\end{equation}

respectively, 
where $p$ is a free parameter varying between 0.5 and 1, 
indicative of how close  our modified model is to the models described with Equations \ref{subida1b} and \ref{bajada1D}. 
In this work, we refer to the diffusion coefficient $D$ when the opposite is not specified. 
The physical meaning of the additional diffusion coefficient $D'$ was studied independently in Sects. \ref{diffd2} and \ref{diffh2}. 

\section{Experimental results}
\label{results}

Two different a-C:H analogs with thicknesses 0.9 $\mu$m and 2.2 $\mu$m were produced under the same conditions in the SICAL-P setup and 
subsequently studied in the ISAC setup. 
Diffusion experiments were carried out at 85 K, 95 K, 105 K, and 110 K for the former sample, and at 95 K, 105 K, 110 K, and 120 K for the 
latter (thinner samples allow the study of diffusion at lower temperatures, since diffusion of molecules through these samples takes less time 
even though the diffusion coefficient is lower, see Sect. \ref{sicalx}). 
In the case of the a-C:D analogs, up to five samples with thicknesses 0.2 $\mu$m, 1.3 $\mu$m, 2.6 $\mu$m, 3.4 $\mu$m, and 5.2 $\mu$m 
were studied in the SICAL-X setup. Experiments were performed at a wide range of temperatures (from 95 K to 170 K). 
As explained in Sect. \ref{isac}, diffusion coefficients for the D$_{2}$ molecules in the a-C:D analogs are lower than those of the 
H$_{2}$ molecules through the a-C:H analogs at the same temperature. 
Therefore, to keep the duration of the experiments within reasonable
limits, experiments with deuterated analogs were carried out at slightly higher temperatures than those performed with hydrogenated analogs of 
similar thickness.

\subsection{IR spectra of the a-C:H and a-C:D analogs}
\label{ir}

\begin{figure}
 \includegraphics[width=9.25cm]{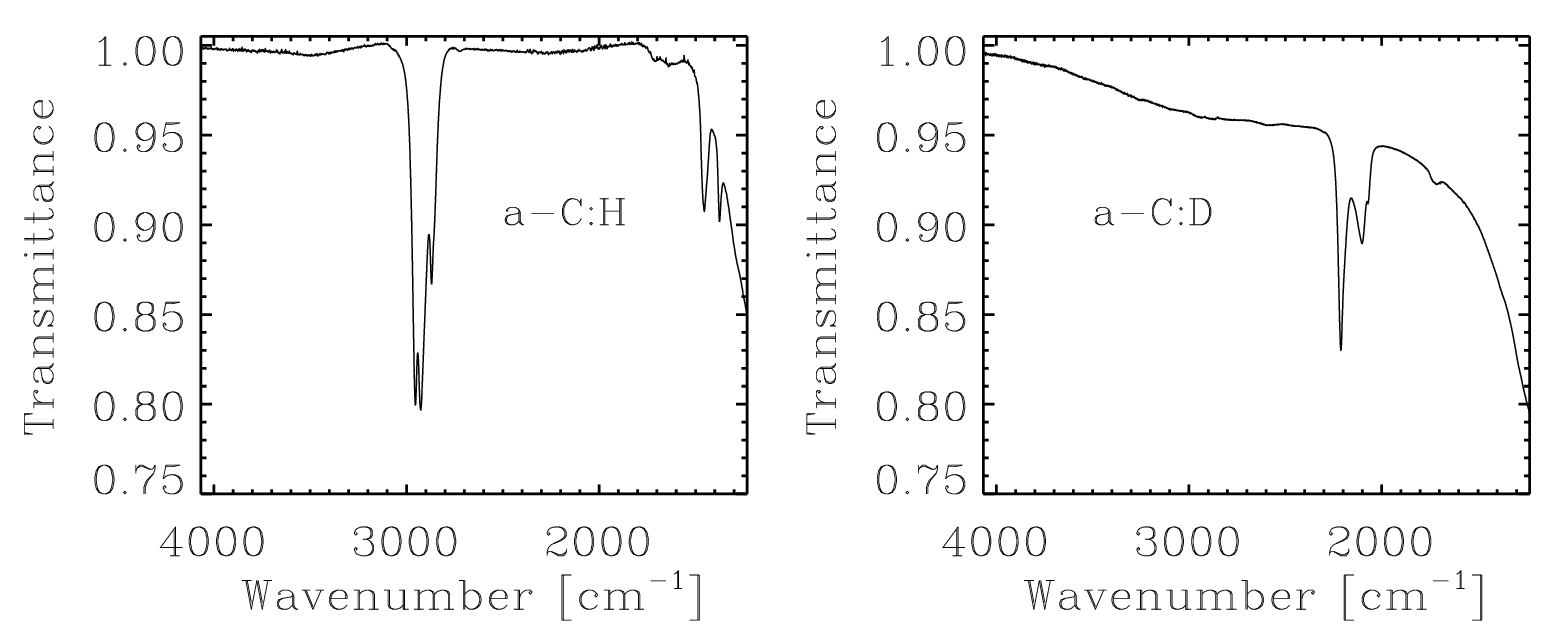}
 \caption{IR transmittance spectrum of an a-C:H analog (\textit{left panel}), and an a-C:D analog (\textit{right panel}) 
 deposited on an MgF$_{2}$ substrate.Spectra were collected at 110 K.}
 \label{IR}
\end{figure}

Figure \ref{IR} shows the mid-IR transmittance spectra of an a-C:H analog (left panel, $l$ = 2.2 $\mu$m), 
and an a-C:D analog (right panel, $l$ = 2.6 $\mu$m), collected at 110 K, between 4000 cm$^{-1}$ and 1300 cm$^{-1}$. 
Transmittance of the MgF$_{2}$ substrate starts decreasing strongly below 1500 cm$^{-1}$. 

As explained in Sect. \ref{intro}, the asymmetric C-H stretching modes corresponding to methyl (-CH$_{3}$) and methylene (-CH$_{2}$-) 
groups led to two absorption peaks at 2955 cm$^{-1}$ and 2925 cm$^{-1}$, respectively, in the left panel of Fig. \ref{IR}. 
The symmetric stretching modes at 2873 cm$^{-1}$ and 2857 cm$^{-1}$, respectively, are blended, which  leads to one absorption peak 
in the red side of the 3.4 $\mu$m absorption band. 
The corresponding bending modes are observed at 1460 cm$^{-1}$ and 1380 cm$^{-1}$ for the methylene and methyl groups, respectively. 
Absorption at $\sim$1600 cm$^{-1}$ is assigned to the C=C stretching mode, which corresponds to the olefinic fraction of the analog. 

The same modes are shifted to lower frequencies in the a-C:D analogs. 
The assymetric stretching modes, located at 2220 cm$^{-1}$ and 2200 cm$^{-1}$, are blended in the right panel of Fig. \ref{IR}, 
as well as the symmetric modes at 2073 cm$^{-1}$ and 2100 cm$^{-1}$. 
The corresponding bending modes are also shifted to lower frequencies, where the absorption of the MgF$_{2}$ substrate prevents their detection. 

Since only a small fraction of the analogs were photoprocessed during the experiments (see Sect. \ref{sicalx}), IR spectra did not change 
after UV irradiation, 
which is a condition to the hypothesis of a constant flux at $x = 0$.

\subsection{QMS measurements of the outgoing H$_{2}$ or D$_{2}$ flux}
As previously explained, the measured ion current of the mass fragments m/z = 2 and m/z = 4 above the background level provided a measure 
of the outgoing H$_{2}$ and D$_{2}$ fluxes through the $x=l$ surface of the hydrogenated and deuterated amorphous carbon analogs, respectively, 
during the experiments. 

\begin{figure}
\includegraphics[width=9.25cm]{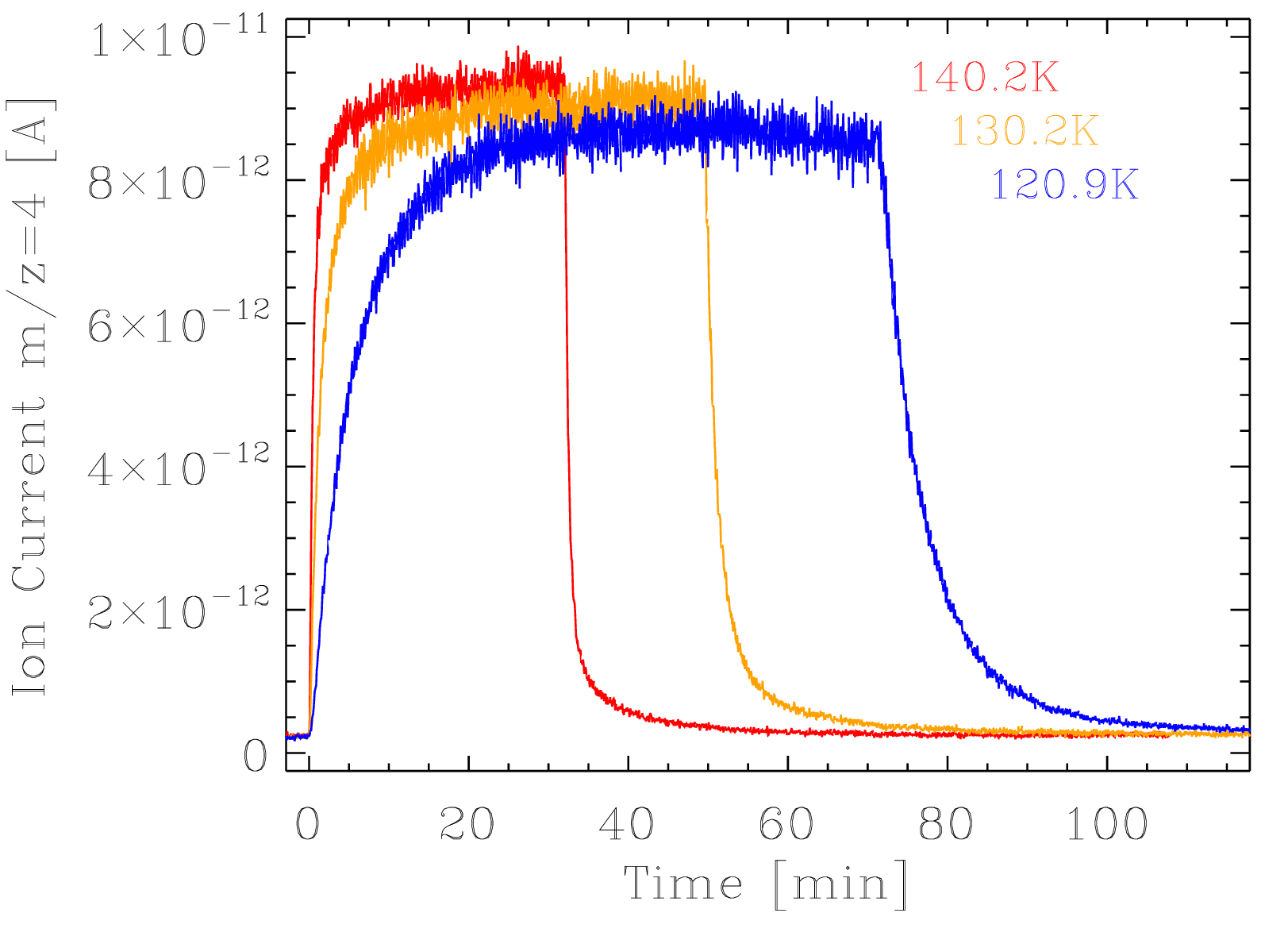}
\caption{Measured m/z = 4 ion current corresponding to the outgoing D$_{2}$ flux during three experiments performed at $\sim$140 K (red solid line), 
$\sim$130 K (yellow solid line), and $\sim$120 K (blue solid line) on a 3.4 $\mu$m thick a-C:D analog. Irradiation starts at $t=0$.} 
\label{irradiationhac}
\end{figure}

\begin{figure}
\includegraphics[width=9.25cm]{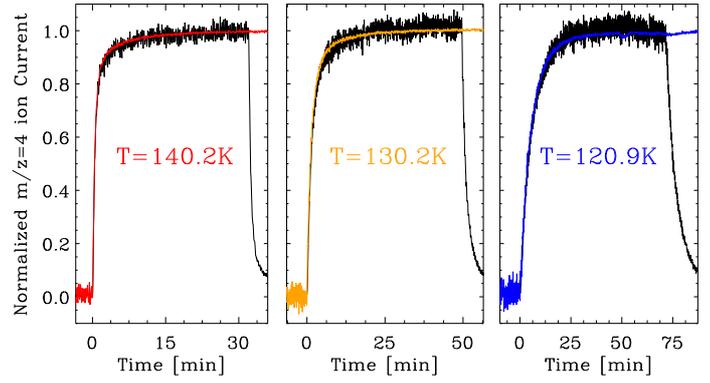}
\caption{Normalized m/z = 4 ion current after background substraction (equivalent to the normalized D$_{2}$ flux) during three experiments 
performed at $\sim$140 K (\textit{left panel}), 
$\sim$130 K (\textit{middle panel}), and $\sim$120 K (\textit{right panel}) on a 3.4 $\mu$m thick a-C:D analog. 
Black solid lines correspond to the normalized D$_{2}$ flux during irradiation (irradiation starts at $t=0$), while 
colored solid lines correspond to 1 - normalized D$_{2}$ flux after irradiation (irradiation stops at $t=0$).}
\label{irradiationhac2}
\end{figure}

Figure \ref{irradiationhac} shows the m/z = 4 ion current measured during three experiments performed at three different temperatures 
with a 3.4 $\mu$m thick a-C:D analog, as an example. 
After onset of the UV irradiation at $t=0$, the outgoing flux of the photo-produced D$_{2}$ molecules increased rapidly at first, 
and then more slowly, until the steady state was reached. 
Diffusion times depended on the diffusion coefficient $D$. 
At higher temperatures, a higher $D$ value led to a faster diffusion, and therefore, to a faster increase of the m/z = 4 ion current, 
while diffusion at lower temperatures was slower. 
The ion current value at the steady state depends not only on the diffusion coefficient (which in turn depends on the temperature), 
but also on the VUV photon flux reaching the sample, which may change slightly from one experiment to another. 
Parallel experiments with different VUV photon fluxes were carried out to check that the derived value of the diffusion coefficient at 
a given temperature did not vary with the value of the UV flux.  
Once the steady state was reached, the shutter was placed in front of the sample, preventing the VUV photons from reaching the analog, 
and thus stopping the D$_{2}$ production. The m/z = 4 ion current then decreased back to the background value, again rapidly at first and more slowly later.  
As for the increase of the outgoing D$_{2}$ flux, decay time also depended on the diffusion coefficient. 
As expected from Equations \ref{subida1b} and \ref{bajada1D}, once the background level was subtracted, 
the normalized ion current (which is equivalent to the normalized D$_{2}$ flux) during irradiation 
was equal to unity minus the normalized ion current during decay of the D$_{2}$ flux (see Fig. \ref{irradiationhac2} as an example). 

Evolution of the normalized ion current (i.e., normalized H$_{2}$ or D$_{2}$ flux through the analog surface $x=l$) 
during and after irradiation is modeled by Equations \ref{subida1b} and \ref{bajada1D}, respectively. 
The value of $D,$ which best fits the measured ion currents during the experiments, would be the diffusion coefficient of H$_{2}$ (D$_{2}$) 
molecules through the a-C:H (a-C:D) analogs at a given temperature. 
Evolution of the diffusion coefficient with the temperature is described by Eq. \ref{diffT}. 

Most of the experiments were carried out with a-C:D analogs in the SICAL-X setup. 
Derived $D$ values for the D$_{2}$ molecules at different temperatures enabled us to calculate the 
pre-exponential factor $D_{0}$ and the activation energy $E_{D}$ for the diffusion of D$_{2}$ through the deuterated analogs. 
Results are presented in Sect. \ref{diffd2}. 
Complementary experiments were performed with two a-C:H analogs in the ISAC setup. 
Calculated $D_{0}$ and $E_{D}$ values are presented in Sect. \ref{diffh2}. 
$E_{D}$ was expected to be similar in both cases. 

\subsubsection{Modeling of the D$_{2}$ diffusion through a-C:D analogs}
\label{diffd2}

\begin{figure*}
\includegraphics[width=18.5cm]{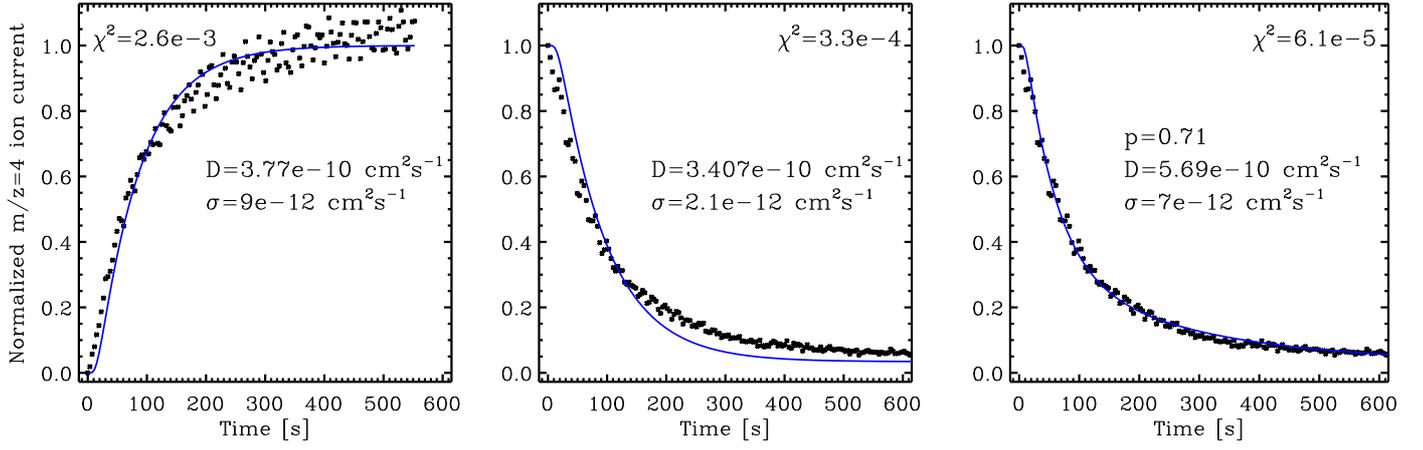}
\caption{Evolution of the normalized m/z = 4 ion current (black dots), equivalent to the normalized D$_{2}$ flux at the $x=l$ surface, 
during (\textit{left panel}) and after (\textit{middle} and \textit{right panels}) VUV irradiation of a 2.6 $\mu$m a-C:D analog at 130 K. 
Theoretical models fitting the experimental data are presented (blue solid lines), 
along with the derived diffusion coefficient $D$ and the 1$\sigma$ associated error. 
The model shown in the \textit{left panel} is described by Eq. \ref{subida1b}. 
The \textit{middle} and \textit{right panels} show the models described by Equations \ref{bajada1Dback} and \ref{bajada2Dback}, respectively. 
The free parameter $p$ of the latter is an indicator of how close the model is to the one described by Eq. \ref{bajada1Dback}. 
The $\chi^{2}$ values associated with the fits of the experimental data during and after irradiation cannot be directly compared since 
they are not normalized.}
\label{ajuste}
\end{figure*}

As explained above, we estimated for every experiment a diffusion coefficient $D$ for the diffusion of D$_{2}$ molecules 
through the a-C:D analogs at  
a given temperature by fitting the models described with Equations \ref{subida1b} and \ref{bajada1D} to the experimental evolution of 
the normalized m/z = 4 ion current during and after the irradiation of the sample, respectively, 
with the linfit procedure programmed with the IDL programming language. 
This procedure finds the free parameter $D$ which minimizes the $\chi^{2}$ parameter.

Left panel of Fig. \ref{ajuste} shows the evolution of the normalized m/z = 4 ion current, i.e., the normalized outgoing D$_{2}$ flux, 
during irradiation of a 2.6 $\mu$m a-C:D analog at 130 K, along with the model described by Eq. \ref{subida1b} that 
best fits the experimental results. 
The diffusion coefficient derived for that experiment is also shown in the figure. 
This value is the same within 10\% to the diffusion coefficient derived from the decay of the normalized outgoing D$_{2}$ flux in the 
same experiment (middle panel of Fig. \ref{ajuste}). 
In this case, instead of subtracting the background level of the m/z = 4 ion current, we included it as a free parameter $b$ in 
Eq. \ref{bajada1D}:

\begin{equation}
F(l,t) = (1 - b) \cdot \frac{4}{\pi} \times \sum_{n = 0}^{\infty}\frac{(-1)^{n}}{(2n + 1)} 
\cdot exp[-\frac{(2n+1)^{2} \pi^{2} D t}{4 l^{2}}] + b.
\label{bajada1Dback}
\end{equation}

We note that $\chi^{2}$ values in the left and middle panels of Fig. \ref{ajuste} cannot be directly compared since they are not normalized. 
However, since there is a larger dispersion in the experimental data collected at the steady state than in the background level, 
we consider the diffusion coefficient derived from the decay curves more accurate, and they have thus been used preferently. 

As explained in Sect. \ref{modelo}, experimental data were better described by a modified model including a second term with an 
additional diffusion coefficient $D'$. 
Adding a second term to Eq. \ref{bajada1Dback} results in   

\begin{equation}
\begin{split}
F(l,t) = & p \cdot \frac{4}{\pi} \times \sum_{n = 0}^{\infty}\frac{(-1)^{n}}{(2n + 1)} 
\cdot exp[-\frac{(2n+1)^{2} \pi^{2} D t}{4 l^{2}}]\\ 
+ & (1-p-b) \cdot \frac{4}{\pi} \times \sum_{n = 0}^{\infty}\frac{(-1)^{n}}{(2n + 1)} 
\cdot exp[-\frac{(2n+1)^{2} \pi^{2} D' t}{4 l^{2}}]\\
 + & b, 
\label{bajada2Dback}
\end{split}
\end{equation}

which reduced the $\chi^{2}$ value of the fit (see right panel of Fig. \ref{ajuste}). 
Estimated diffusion coefficients $D$ with the modified model were of the same order as those found with the model described by Eq. \ref{bajada1Dback}. 
Values of the free parameter $p$ usually varied between $\sim$0.7 and $\sim$0.9. 
The physical meaning of the $D'$ diffusion coefficient was studied independently (see below). 

\begin{figure}
\includegraphics[width=9.25cm]{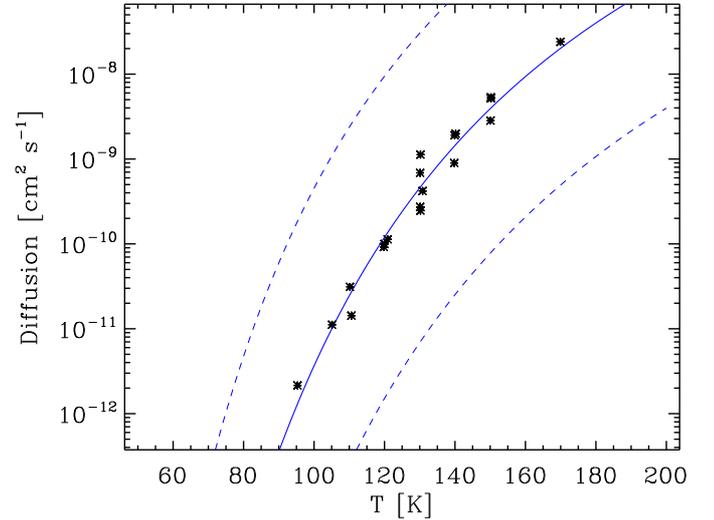}
\caption{Evolution of the experimental diffusion coefficient $D$ for the diffusion of D$_{2}$ molecules through a-C:D analogs 
with the temperature (black circles), along with the model described by 
Eq. \ref{diffT} that best fits the experimental data (blue solid line), and the associated 3$\sigma$ limits (blue dashed lines).}
\label{ajusteT}
\end{figure}

The estimated 1$\sigma$ errors presented in Fig. \ref{ajuste} for the diffusion coefficient $D$ are model-dominated. 
We note that diffusion coefficients derived with different experiments at the same temperature can vary up to a factor $\sim$6 
(see Fig. \ref{ajusteT}). 
This was taken into account for the $D_{0}$ and $E_{D}$ estimation. 


A total of 19 experiments at temperatures between 95 K and 170 K were carried out with five a-C:D analogs. 
The diffusion coefficient at intermediate (120 K - 140 K) temperatures was probed with most of the analogs, 
while $D$ value at low (high) temperatures was probed only with thin (thick) analogs. 
As explained in Sect. \ref{intro}, dependence of the diffusion coefficient $D$ with the temperature follows an Arrhenius-type equation. 
Equation \ref{diffT} changes into a line by applying natural logarithm to both sides of the equation 

\begin{equation}
ln[D] = ln[D_{0}] - E_{D} \cdot \frac{1}{T},  
\label{diffTlin}
\end{equation}

where the slope $E_{D}$ corresponds to the activation energy of the diffusion process. 
Fitting Eq. \ref{diffTlin} to the derived diffusion coefficients $D$ from the experiments with the linfit procedure programmed with 
the IDL programming language led to an activation energy of $E_{D}$ = 2090 $\pm$ 90 K 
for the diffusion of D$_{2}$ molecules through the a-C:D analogs, and 
$D_{0}$ = 0.0045$^{+ 0.0050}_{- 0.0023}$ cm$^{2}$ s$^{-1}$ 
for the pre-exponential factor. 
This activation energy falls in the same range as that found in \citet{falconneche01} for the diffusion of rare gases and small molecules through 
carbonaceous polymer materials.  
Figure \ref{ajusteT} shows the evolution of the experimental and modeled diffusion coefficient with the temperature, along with the associated 
3$\sigma$ limits. 

\begin{figure}
\includegraphics[width=9.25cm]{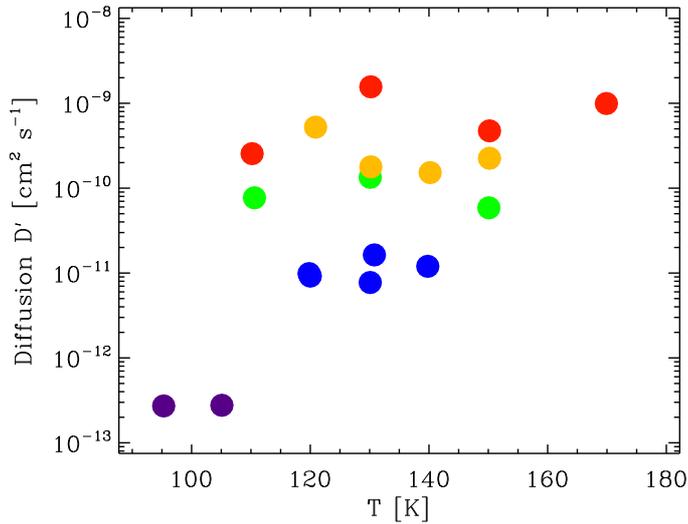}
\caption{Additional diffusion coefficient $D'$ derived for the experiments with a-C:D analogs of thickness 5.2 $\mu$m (red circles), 
3.4 $\mu$m (yellow 
circles), 2.6 $\mu$m (green circles), 1.3 $\mu$m (blue circles), and 0.2 $\mu$m (purple circles), using the model described by 
Eq. \ref{bajada2Dback}.}
\label{diff2}
\end{figure}

On the other hand, the additional diffusion coefficient $D'$ introduced to improve the model 
did not depend on the temperature, but on the thickness of the analogs, as seen in Fig. \ref{diff2}. 
The second term in Equations \ref{subida2D} and \ref{bajada2D} could be an artifact that accounts for the differences between the model 
described by Equations \ref{subida1b} and \ref{bajada1D}, and the real process taking place during the experiments. 
For example, the entering D$_{2}$ flux established during irradiation is assumed to enter the sample at the surface $x = 0$ with 
infinitesimal thickness, while in the real process a finite (yet negligible) thickness of the sample is processed by the VUV photons. 
In which case, the diffusion coefficient $D'$ would have no physical meaning.   
Alternatively, this second term could account for a parallel diffusion process taking place along with the "main" studied diffusion 
represented by the coefficient $D$. 
 In that case, the diffusion coefficient $D'$ would describe this parallel diffusion, which could be taking place, for example, 
through the pores or cracks of the samples. 
Since $D'$ increases with the analogs' thickness (see Fig. \ref{diff2}), pores or cracks should be larger in thicker plasma-produced analogs. 


\subsubsection{Modelling of the H$_{2}$ diffusion through a-C:H analogs}
\label{diffh2}

\begin{figure}
\includegraphics[width=9.25cm]{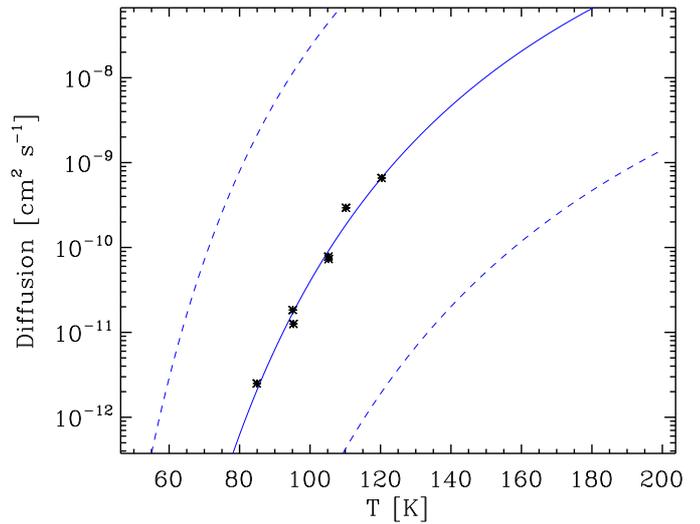}
\caption{Evolution of the experimental diffusion coefficient $D$ for the diffusion of H$_{2}$ molecules through a-C:H analogs 
with the temperature (black circles), along with the model described by 
Eq. \ref{diffT} that best fits the experimental data (blue solid line), and the associated 3$\sigma$ limits (blue dashed lines).}
\label{ajusteTH}
\end{figure}

\begin{figure}
\includegraphics[width=9.25cm]{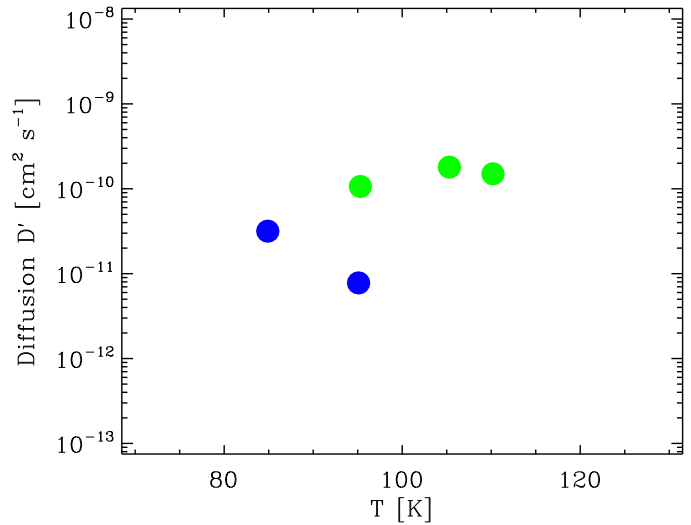}
\caption{Additional diffusion coefficient $D'$ derived for the experiments with a-C:H analogs of thickness 2.2 $\mu$m (green circles), 
and 0.9 $\mu$m (blue circles), using the model described by Eq. \ref{bajada2Dback}.}
\label{diff2H}
\end{figure}

A total of seven complementary experiments were carried out with 2 a-C:H analogs, at temperatures between 85 K and 120 K. 
The modified model described by Eq. \ref{bajada2Dback} was also used to fit the normalized m/z = 2 ion current measured during the experiments. 
Estimated diffusion coefficients $D$ for the diffusion of H$_{2}$ molecules through the a-C:H analogs were found to be approximately one order 
of magnitude higher than those of D$_{2}$ molecules through a-C:D analogs measured at the same temperatures. 
However, dependence of the diffusion coefficient with temperature was similar in both cases. 
The estimated activation energy and diffusion at infinite temperature were $E_{D}$ = 1660 $\pm$ 110 K, and 
$D_{0}$ = 0.0007$^{+ 0.0013}_{- 0.0004}$ cm$^{2}$ s$^{-1}$, respectively (see Fig. \ref{ajusteTH}). 
The errors could be slightly larger due to the assumptions made for the m/z = 2 background level changes when switching the VUV lamp on and off, 
since the metallic shutter used to block the VUV photons in the SICAL-X setup after irradiation was not available in the ISAC setup. 
The additional diffusion coefficient in Eq. \ref{bajada2Dback} followed the same behavior as in Sect. \ref{diffd2}, as shown in Fig. \ref{diff2H}, 
with averaged values for similar thicknesses higher by a factor of $\sim$2. 

\section{Astrophysical implications}
\label{astro}
The mobility and desorption of H$_{2}$ and D$_{2}$ on water ice, mineral, or graphite surfaces are high 
(see, e.g., Amiaud et al. 2015, Acharyya 2014,  Vidali \& Li 2010, Fillion et al. 2009, Haas et al. 2009, and references therein),  
with typical activation energies of only a few hundreds of K. 
This ensures the rapid molecular hydrogen desorption in high dust grain temperature regions, such as some PDRs, 
but at the same time it constitutes an issue when trying to explain the formation of H$_2$ molecules by the surface recombination of H atoms, 
since they may not stay physisorbed for long enough  on the surface to recombine. 
Alternatively to these formation paths, 
a-C:Hs are energetically processed in the ISM, leading to the destruction of the aliphatic C-H component (which is not detected in dense regions), 
and the formation of hydrogen molecules. 
Recent models help in explaining the H$_{2}$ formation at such apparent high temperatures, because of the time fraction spent by small 
grains at low temperature between transient heating events \citep{bron16}. 
The subsequently formed hydrogen molecules diffuse out of the carbonaceous particles and contribute to the total H$_{2}$ abundance in the ISM 
\citep{alata14}.  
In the diffuse ISM, re-hydrogenation of the amorphous carbon particles by atomic H equilibrates the destruction of the aliphatic C-H component 
by UV photons (mainly) and cosmic rays. 
In the dense ISM, re-hydrogenation is much less efficient. 
However, the interstellar UV field cannot penetrate dense clouds, and destruction only by 
cosmic rays (directly or indirectly through the generated secondary UV field) cannot account for the disappearance of the 3.4 $\mu$m absorption 
band. 
\citet{godard11} state that this dehydrogenation should therefore take place in intermediate regions such as translucent clouds or PDRs.

In this scenario, when H$_{2}$ molecules are produced in the bulk by photolytic reactions, they diffuse out slowly outward of the grains 
contrary to the production and immediate release of H$_2$ that are formed on the surface of interstellar solids. 
Surface Eley-Rideal or Langmuir–Hinshelwood mechanism 
\citep[see, e.g., ][]{roser03,hornekaer03,islam07,vidali07,latimer2008,cazaux2008,mennella08,vidali09,lemaire10,sizun2010,vidali13,hama13,bron2014,amiaud2015}
led to the H$_{2}$ molecule desorption with debated amounts of internal excitation energy. 
By contrast, the diffusion process in the bulk allows the molecules to thermalize during their escape path toward the surface.  
One thus expects no high vibrational level excitation for these molecules contrary to the possible hydrogen surface recombination processes.

Formation of hydrogen molecules and subsequent diffusion in the bulk of the a-C:H particles also enables us to thermalize the ortho-to-para (OPR) 
ratio to the dust temperature, which is lower than the gas temperature in PDRs. 
For dust temperatures of 55-70 K typical of a warm PDR such as the Orion Bar nebula \citep[see, e.g., ][and references therein]{guzman11}, 
an OPR of $\sim$1 is expected. 
Observed OPR values in PDRs are indeed around 1, lower than the value of $\sim$3 expected from the excitation temperature of the H$_{2}$ 
rotational lines \citep{fuente99,habart03,habart11}. 
Alternatively, \citet{bron16} propose the ortho-to-para conversion of physisorbed H$_{2}$ molecules on dust grain surfaces to explain these low OPR 
values. 
This process is only efficient on cold dust grains and, therefore, dust temperature fluctuations need to be invoked in that case. 

Diffusion of H$_{2}$ through the a-C:H particles is characterized by the diffusion coefficient $D$, which depends on the dust temperature 
according to Eq. \ref{diffT}. 
With the $D_{0}$ and $E_{D}$ values estimated in Sect. \ref{diffd2} 
we can extrapolate $D$ values to the temperature of a PDR region, and obtain a typical decay-time constant for the 
release of the hydrogen molecules from the a-C:Hs in space.  
Interstellar a-C:Hs can be approximated to a slab of thickness $2r$,  $r$ being the typical radius of a carbonaceous dust particle.  
If the a-C:H is initially filled with a concentration  $C_{0}$ of H$_{2}$ molecules, and the molecules are released to the gas phase through the 
surfaces $x = -r$ and $x = r$ maintained at zero concentration, then the average hydrogen concentration in the slab at time $t$ is 
given by

\begin{equation}
C_{av}(t) = \frac{4 C_{0}}{\pi^{2}} \times \sum_{n = 0}^{\infty}\frac{1}{(2n + 1)^{2}} 
\cdot exp[-\frac{(2n+1)^{2} \pi^{2} D t}{4 r^{2}}], 
\label{slab}
\end{equation}

according to \citet{carslaw}. 
We define the decay-time constant $\tau$ as the time interval needed for the average hydrogen concentration in the slab to be 10\% of the initial 
value (i.e., $\frac{C_{av}(\tau)}{C_{0}}=0.1$). 
Adopting a typical $\sim$0.1 $\mu$m grain radius,  
we calculated $\tau$ for two different temperatures: 30 K, 
representative of a cold PDR such as the Horsehead nebula; and 55 K, the lower limit of a warmer PDR such as the Orion Bar nebula 
\citep[see, e.g., ][and references therein]{guzman11}. 
While at 55 K calculated $\tau_{55}$ is about 19 years, at 30 K $\tau_{30}$ would be about 1.1 $\times$ 10$^{15}$ years, since 
the diffusion coefficient depends strongly on the temperature. 
If we instead adopt  half the typical radius for a dust grain ($r$ $\sim$ 0.05 $\mu$m), decay times are reduced to a quarter (see Eq. \ref{slab}), 
leading to $\tau_{55}$ $\sim$ 5 years and $\tau_{30}$ $\sim$ 3 $\times$ 10$^{14}$ years. 
The latter decay-time is way longer than the typical dynamical time of a PDR \citep[see, e.g., ][]{goldsmith07,glover07}. 
To decrease the desorption time of the photo-produced hydrogen molecules in the coldest interstellar regions, several solutions can be invoked. 
On one hand, transient heating episodes of the a-C:H particles by, for example, cosmic rays could increase the dust temperature 
(and therefore the diffusion of the H$_{2}$ molecules), thus reducing significantly the release time. 
On the other hand, the presence of open channels in the grains, or a higher surface-to-volume ratio than that of the compact films used in the 
laboratory simulations could also assist the hydrogen release from the a-C:H particles. 

\section{Conclusions}
\label{conclusiones}
We have explored the diffusion of photo-produced H$_{2}$ (D$_{2}$) molecules through a-C:H (a-C:D) analogs. 
Hydrogenated amorphous carbon particles (which harbor between 5\% and 30\% of the total C cosmic abundance) 
are energetically processed in the ISM, leading to the loss of the aliphatic C-H component and the 
formation of hydrogen molecules that diffuse out of the particles, contributing to the total H$_{2}$ abundance. 
This constitutes an alternative additional formation pathway to the hydrogen surface recombination process 
that allows the hydrogen molecules to thermalize to the dust temperature before passing into the gas phase,  
leading to no high vibrational level excitation, and to OPR values similar to those observed in PDRs. 

We have simulated this process in the laboratory using plasma-produced a-C:H and a-C:D analogs.   
The surface of the analogs in contact with the substrate was irradiated by VUV photons under astrophysically relevant conditions.  
Photo-produced H$_{2}$ and D$_{2}$ molecules subsequently diffused through the analogs, eventually reaching the opposite surface, and 
passing into the gas phase. 
Molecules released from the analogs were detected by a QMS. 
The measured m/z = 2 and m/z = 4 ion current corresponded to the outgoing H$_{2}$ or D$_{2}$ flux, respectively,  
which was compared to the expected flux from the diffusion model that best fitted the experimental measurements, 
enabling us to derive a diffusion coefficient that described the diffusion process. 
A modified diffusion model with two different diffusion coefficients, $D$ and $D'$ was used. 
The diffusion coefficient $D$ described the diffusion of the molecules through the HAC material, which depended on the temperature of the 
analogs, following an Arrhenius-type equation. 
Experiments at several temperatures were carried out, estimating a diffusion coefficient $D$ for every experiment.  
This allowed us to derive an activation energy $E_{D}$ of the diffusion process.  
Estimated $E_{D}$ was 1660 $\pm$ 110 K for the diffusion of H$_{2}$ through the a-C:H analogs, 
and 2090 $\pm$ 90 K  for the diffusion of D$_{2}$ through the a-C:D analogs. 
The pre-exponential factors were also derived 
($D_{0}$(H$_{2}$) = 0.0007$^{+0.0013}_{-0.0004}$ cm$^{2}$ s$^{-1}$, and $D_{0}$(D$_{2}$) = 0.0045$^{+0.005}_{-0.0023}$ cm$^{2}$ s$^{-1}$). 
%
The additional diffusion coefficient $D'$ did not depend on the temperature, but on the thickness of the analogs. 
This coefficient could 
trace 
the differences between the model and the real process taking place in the laboratory, 
or, alternatively, it could be accounting for a parallel diffusion process taking place, for example, through the pores or cracks 
of the analogs.  

Using these experimental values, we extrapolated the diffusion coefficient $D$ to two different temperatures representative of PDR regions, 
where the destruction of the C-H bonds and formation of H$_{2}$ molecules is expected to take place. 
A typical decay-time constant $\tau$ was calculated characterizing the release of the H$_{2}$ molecules from the a-C:H particles. 
Transient heating episodes of the dust particles or other alternative solutions need to be invoked for the release of the hydrogen molecules in 
cold regions where the typical diffusion times exceed the dynamical time of these regions.

\begin{acknowledgements}
This research has been financed by the ANR and French INSU-CNRS program Physique et Chimie du Milieu Interstellaire (PCMI), 
and by the Spanish MINECO under projects AYA-2011-29375 and AYA2014-60585-P. R.M.D. benefited from a FPI grant from Spanish MINECO. 
The authors acknowledge funding support from the PICS (Projet International de Coopération Scientifique) between the CNRS and CSIC 
which consolidated this French and Spanish teams' cooperation. 
We also thank the anonymous reviewer for  constructive remarks

\end{acknowledgements}

\end{document}